\documentstyle[twoside,fleqn,espcrc2]{article}

\def\3{\ss}

\newcommand{\DD}{\not\!\!{D}}
\newcommand{\p}{\not\!{\partial}}
\newcommand{\A}{\not\!\!{A}}
\newcommand{\half}{\mbox{\small $\frac{1}{2}$}}
\newcommand{\sze}{\scriptsize}
\newcommand{\Z}{\mbox{\scriptsize \bf Z}}
\newcommand{\AmS}{{\protect\the\textfont2
  A\kern-.1667em\lower.5ex\hbox{M}\kern-.125emS}}

\title{A method for putting chiral fermions on the lattice}

\author{M. G\"ockeler%
\address{Institut f\"ur Theoretische Physik, RWTH Aachen,
        Sommerfeldstra\3e, W-5100 Aachen, Germany}$\! \! ^,$%
\address{H\"ochstleistungsrechenzentrum HLRZ,
         c/o KFA, Postfach 1913, W-5170 J\"ulich, Germany}%
        \thanks{Work supported by the
                Deut\-sche For\-schungs\-ge\-mein\-schaft.}
        and
        G. Schierholz$^{\rm b,}$%
        \address{Deutsches Elektronen-Synchrotron DESY,
        Notkestra\3e 85, W-2000 Hamburg 52, Germany}}

\begin{document}

\begin{abstract}
We describe a method to put chiral gauge theories on the lattice.
Our method makes heavy use of the effective action for
chiral fermions in the
continuum, which is in general complex. As an example we
discuss the chiral Schwinger model.
\end{abstract}

\maketitle

\section{INTRODUCTION}

In spite of the well-known importance of chiral gauge theories there
is very little nonperturbative information available about such models.
Standard nonperturbative techniques seem to fail, in particular the
lattice approach faces great difficulties due to the notorious
doubling problem. To overcome these difficulties it has been suggested
to keep the fermionic degrees of freedom in the continuum and to
latticize only the gauge fields~\cite{confer}. However, the details
of such a procedure have not yet been worked out. In the present note
we want to outline a few steps along this line (see also~\cite{rome}).

\section{EFFECTIVE ACTION}

First of all we need the effective action for chiral fermions in
the background of a continuum gauge field. This has been discussed
by several groups of authors~\cite{alva,ball,niemi}. We
shall mainly follow the approach of
Alvarez-Gaum\'e et al.~\cite{alva}.

Working in euclidean space we choose hermitian $\gamma$-matrices,
\begin{equation}
 \gamma_\mu = \gamma_\mu^+ \,, \quad
   \{ \gamma_\mu, \gamma_\nu \} = \delta_{\mu,\nu} \,,
\end{equation}
and define $\gamma_5$ such that
\begin{equation}
   \{ \gamma_5, \gamma_\mu \} = 0 \,,
   \gamma_5^2 = 1 \,,
   \gamma_5^+ = \gamma_5 \,.
\end{equation}
So the chiral projectors are given by
\begin{equation}
  P_\pm = \frac{1}{2} \left( 1 \pm \gamma_5 \right) \,.
\end{equation}
The gauge fields are taken to be antihermitian:
 \begin{equation}
   A_\mu^+ (x) = - A_\mu (x)\,.
 \end{equation}
We write the chiral Dirac operator as
\begin{equation}
  \mbox{i} \hat{D} (A) = \mbox{i} \left(\p + \A P_+ \right) \,.
\end{equation}
Consequently, both chiralities of fermions are present, but the
gauge field couples only to one of them. For the vector Dirac
operator we use the standard notation
\begin{equation}
  \mbox{i} \DD (A) = \mbox{i} \left(\p + \A \right) \,.
\end{equation}

Formally, i.e.\ up to regularization, the effective action
$W[A]$ can now be expressed in terms of the functional integral
 \begin{eqnarray}
 & & \mbox{e} ^{-W[A]}   \nonumber \\
 & &    = \int \! {\cal D} \psi {\cal D} \bar{\psi}
    \exp \left( - \int \! d^4 \! x \bar{\psi} (x) \mbox{i} \hat{D} (A)
      \psi (x) \right )  \nonumber \\
 & &    =  \, \det \left( \mbox{i} \hat{D} (A) \right) \,.
 \end{eqnarray}
Since i$\hat{D} (A)$ is not hermitian, $W[A]$ is expected to be
complex in general. Up to local counterterms, the real part is
given by the effective action for a Dirac fermion interacting
with the gauge field $A$,
\begin{equation}
  \mbox{Re} W[A] = - \half \ln \det ( \mbox{i} \DD (A)) \,,
\end{equation}
and so it is gauge invariant.

For the imaginary part one obtains
 \begin{eqnarray}
 & & \mbox{Im} W[A]   \nonumber \\
 & & = \frac{1}{2 \mbox{i}} \int_0^1 \! d\tau \, \mbox{Tr}
    \gamma_5 \left( \frac{d}{d \tau} \mbox{i} \DD(A_\tau) \right)
      \left( \mbox{i} \DD(A_\tau) \right) ^{-1}
 \nonumber \\
 & &  + \, \frac{1}{\mbox{i}} \int_0^1 \! d\tau \, \mbox{Tr}
  \left( \frac{\partial A_\tau^\mu}{\partial \tau}
  J_\mu (A_\tau) \right) \,.
  \label{ImW} \end{eqnarray}
Here Tr denotes the operator trace, $A_\tau$ interpolates
between $A_0=0$ and $A_1 = A$ (e.g.\ $A_\tau = \tau A$)
and the current $J_\mu$ accounts for the difference between
the covariant and the consistent anomaly.
Since we are ultimately only interested in anomaly free models,
where contributions from these currents cancel, we shall omit the
corresponding terms in the following.
(Strictly speaking, the above formula can be derived only for
perturbative gauge field configurations, i.e.\ configurations such
that the topological charge vanishes and the Dirac operator has
no zero modes. However, one could more generally interpolate
between two configurations $A_0$ and $A_1$ to obtain
$\mbox{Im} W[A_1] - \mbox{Im} W[A_0]$.)
Note that in eq.~(\ref{ImW}) only the {\em vector} Dirac operator
appears.

We can only mention in passing that $\mbox{Im} W[A]$ may be
expressed in terms of a topological quantity, the spectral
asymmetry or $\eta$-invariant of a vectorlike Dirac operator
in five dimensions (for a review, see \cite{ballr}).

\section{METHOD OF SIMULATION}

Since the effective action governing the dynamics of the gauge
fields has turned out to be complex, one runs into well-known
problems when attempting numerical simulations. At least in
principle, however, one could incorporate $\mbox{Im} W[A]$
in the observable and update the gauge fields with the action
\begin{equation}
  S_G + \mbox{Re} W
  = S_G - \half \ln \det ( \mbox{i} \DD ) \,,
\end{equation}
where $S_G$ denotes the pure gauge field action. Remember that
$\mbox{Re} W$ is given by a vectorlike fermion action, whose
lattice version should be unproblematic.

Irrespective of the details of the simulation method one
has to solve the question of how to calculate $\mbox{Im} W$
for a given lattice gauge field configuration. A naive procedure
would be to replace $\DD$ in eq.~(\ref{ImW}) by some lattice
version of the Dirac operator, e.g.\ Wilson fermions. However,
since $\mbox{Im} W$ is related to a topological
quantity, the $\eta$-invariant, a look at the topological
charge might be advisable. In that case such a procedure does
not work: An expression for the lattice topological charge
constructed out of a few plaquettes only suffers from large
perturbative contributions leading to a divergent topological
susceptibility in the continuum limit. The geometrical
constructions of the lattice topological charge, on the other
hand, which avoid this problem, can be viewed as computing
the continuum charge from a continuum gauge field derived from
the lattice gauge field by a suitable interpolation
\cite{lusch,congau}. Therefore we propose to define
$\mbox{Im} W$ by $\mbox{Im} W[A]$ evaluated
for this interpolated gauge potential $A$. A suitable
interpolation has been explicitly given in \cite{congau},
although the construction
is not completely trivial as several constraints have to be
fulfilled. In particular, gauge covariance has
to be guaranteed.

In order to really calculate $\mbox{Im} W[A]$
one has to introduce some regularization. From the numerical
point of view it seems to be most convenient to use the lattice
again, i.e.\ to compute on finer and finer sublattices of the
original lattice anticipating that the results will converge.
(The rigorous analytical investigations apply other regularization
schemes, e.g.\ Pauli-Villars.)

So one would proceed as follows. Start from a lattice gauge field
$U(x,\mu)$ on the original lattice, whose spacing is put equal
to one. Construct from $U$ an interpolated continuum gauge field
$A$. This is then used to calculate parallel transporters
$\tilde{U}(x,\mu)$ on a finer lattice of spacing $\epsilon < 1$.
These link matrices are
raised to the power $\tau$ to arrive at a lattice version of the
gauge potential $A_\tau$. (An alternative consists in computing
the parallel transporters on the finer lattice immediately
from $A_\tau$.) As a regularized version of $\DD (A_\tau)^{-1}$
we can then take the propagator of Wilson fermions in the gauge
field $\tilde{U}^\tau$ given by
\begin{equation}
  G(\tilde{U}^\tau | x,y) = \epsilon^{-8}
  M^{-1}(\tilde{U}^\tau | x,y)
\end{equation}
with the fermion matrix
 \begin{eqnarray}
 & & M(U|x,y)         \nonumber \\
 & & = \, \epsilon^{-4} \biggl \{ \frac{1}{2 \epsilon}
 \sum_\mu (\gamma_\mu - r) U(x,\mu) \delta_{y,x+\hat{\mu}}
 \nonumber \\
 & & - \, \frac{1}{2 \epsilon}
 \sum_\mu (\gamma_\mu + r) U^+(y,\mu) \delta_{y,x-\hat{\mu}}
 \nonumber \\
 & & + \, \frac{4r}{\epsilon} \delta_{x,y} \biggr \} \,.
 \end{eqnarray}
Finally one has to calculate
 \begin{eqnarray}
 & & \frac{1}{2 \mbox{i}} \int_0^1 \! d \tau \epsilon^8 \sum_{x,y}
 \mbox{tr} \, \gamma_5 \left( \frac{d}{d \tau}
  M(\tilde{U}^\tau | x,y)  \right)
 \nonumber \\
 & & \times \, G(\tilde{U}^\tau | y,x)
 \nonumber \\
 & & = \, \frac{1}{2 \mbox{i}} \int_0^1 \! d \tau \epsilon^4 \sum_{x,\mu}
 \biggl \{ \mbox{tr} \, \gamma_5 \frac{1}{2 \epsilon} (\gamma_\mu -r)
 \nonumber \\
 & & \times \, \ln \tilde{U}(x,\mu)  \tilde{U}(x,\mu)^\tau
 G(\tilde{U}^\tau | x+\hat{\mu},x)
 \nonumber \\
 & &  + \, \mbox{tr} \, \gamma_5 \frac{1}{2 \epsilon} (\gamma_\mu +r)
 \ln \tilde{U}(x,\mu)
 \nonumber \\
 & & \times \, \tilde{U}(x,\mu)^{-\tau}
 G(\tilde{U}^\tau | x,x+\hat{\mu})   \biggr \}
 \end{eqnarray}
in the limit $\epsilon \to 0$, where tr denotes the trace over
Dirac and internal indices.

\section{THE CHIRAL SCHWINGER MODEL}

As an example consider the chiral Schwinger model, i.e.\ a
chiral U(1) model in two dimensions. On a square of side length
$L$ the fermionic part of the continuum action reads
\begin{equation}
  \int_0^L \! d^2 \! x \bar{\psi} (x) \left( \mbox{i} \p
   - \A (x) P_+ \right) \psi (x) \,,
\end{equation}
where now $\gamma_5 = \mbox{i} \gamma_1 \gamma_2$ and the potential
$A_\mu (x)$ is real. We assume periodic boundary conditions for
the gauge field, whereas the fermion field is taken antiperiodic.
Choosing the linear interpolation $A_\tau = \tau A$ so that
\begin{equation}
  \mbox{i} \DD (A_\tau) = \mbox{i} \p - \tau \A (x)
\end{equation}
we get for the relevant contribution to $\mbox{Im} W$
(up to regularization)
 \begin{eqnarray}
 & &  \frac{1}{2 \mbox{i}} \int_0^1 \! d\tau \, \mbox{Tr} \,
    \gamma_5 \left( \frac{d}{d \tau} \mbox{i} \DD(A_\tau) \right)
      \left( \mbox{i} \DD(A_\tau) \right) ^{-1}
  \nonumber \\
  & & = \frac{\mbox{i}}{2} \int_0^1 \! d\tau \, \mbox{Tr} \,
    \gamma_5 \A
      \left( \mbox{i} \DD(A_\tau) \right) ^{-1}
  \nonumber \\
  & & = \frac{1}{2} \int_0^1 \! d\tau \int_0^L  \! d^2 \! x
  \, \mbox{tr} \,   \gamma_5 \A (x)  G(A_\tau | x,x) \,,
 \label{ImW2} \end{eqnarray}
where the propagator $G(A|x,y)$ is determined by the equation
\begin{equation}
  \DD_x (A) G(A|x,y) = \delta (x-y) \,.
\end{equation}

Since in two dimensions an explicit expression for
$G(A|x,y)$ is available \cite{schwinger},
we can evaluate (\ref{ImW2}) further.
Being only interested in the gauge invariant contributions we
fix the Landau gauge,
\begin{equation}
  \partial_\mu A_\mu (x) = 0.
\end{equation}
The zero momentum mode of the gauge field requires a special
treatment, so we define
\begin{equation}
  a_\mu := \frac{1}{L^2} \int_0^L \! d^2 \! x A_\mu (x)
\end{equation}
and assume $ - \pi/L < a_\mu < \pi/L $. Introducing point
splitting we find
 \begin{eqnarray}
 & & \int_0^L \! d^2 \! x \, \mbox{tr} \, \gamma_5 \A (x)
   G(A_\tau | x,x+\delta)
  \nonumber \\
 & &  \times \, \exp \left \{ - \mbox{i} \int_x^{x+\delta} \! dz_\mu
     A_\mu (z) \tau \right \}
  \nonumber \\
 & & = \, \frac{L}{\pi} \sum_{n \in \Z^2,n \neq 0}
    (a_2 n_1 - a_1 n_2)/ n^2
  \nonumber \\
 & &  \times \, \sin (L(b+\tau a) \cdot n) \, \mbox{e} ^{- n^2 L^2 /4}
  \nonumber \\
 & & - \, \frac{4 \pi}{L} \sum_{n \in \Z^2}
  \left( a_1 n_2 + \half a_1 - a_2 n_1 - \half a_2 \right)
  \nonumber \\
 & & \times \, \mbox{e}^{-(2 \pi n /L + b + \tau a)^2} /
        (2 \pi n /L + b + \tau a)^2
  \nonumber \\
 & & + \, \mbox{i} L^2 (a_1 \delta_2 - a_2 \delta_1) /
        (\pi \delta^2)
  + {\cal O} (|\delta|)  \,,
 \end{eqnarray}
where $b_1 = b_2 = \pi/L$. Since the divergent term is odd in
$\delta$, symmetric point splitting leads to a finite limit as
$\delta \to 0$. Only the zero momentum mode of the gauge field
contributes in Landau gauge.
Note that working in infinite volume one does not have this
mode and hence finds $\mbox{Im} W = 0$ in
Landau gauge (see, e.g., \cite{rome}).

Which results does our lattice recipe produce in this case?
Unfortunately, an explicit expression for the lattice fermion
propagator in a general background is not available. So we consider
the special case of the gauge field configuration
\begin{equation}
  U(x,\mu) = \mbox{e}^{\mbox{\sze i} a_\mu}
  \,, \quad  -\pi/L < a_\mu < \pi/L \,,
\end{equation}
on the original lattice with lattice spacing 1. It should not
come as a surprise that the interpolated continuum gauge field
turns out to be $A_\mu (x) = a_\mu$. So we find for the parallel
transporters on the sublattice of spacing $\epsilon$
\begin{equation}
  \tilde{U}(x,\mu) = \mbox{e}^{\mbox{\sze i} \epsilon a_\mu}
\end{equation}
and consequently
\begin{equation}
  \ln \tilde{U}(x,\mu) =  \mbox{i} \epsilon a_\mu \,, \quad
  \tilde{U}(x,\mu) ^\tau = \mbox{e}^{\mbox{\sze i}
  \tau \epsilon a_\mu}\,.
\end{equation}
The required Wilson fermion propagator can easily be calculated.
According to our proposal we then have to compute
 \begin{eqnarray}
 & & - \mbox{i} \epsilon^2 \sum_{x,\mu}
 \Bigl \{ \mbox{tr} \, \gamma_5 \frac{1}{2 \epsilon} (\gamma_\mu -r)
  \ln \tilde{U}(x,\mu)
 \nonumber \\
 & & \times \, \tilde{U}(x,\mu)^\tau
 G(\tilde{U}^\tau | x+\hat{\mu},x)
 \nonumber \\
 & & + \, \mbox{tr} \, \gamma_5 \frac{1}{2 \epsilon} (\gamma_\mu +r)
 \ln \tilde{U}(x,\mu)
 \nonumber \\
 & &  \times \, \tilde{U}(x,\mu)^{-\tau}
 G(\tilde{U}^\tau | x,x+\hat{\mu})   \Bigr \}
 \nonumber \\
  & &  = 2 \epsilon \! \sum_k \left[ a_2 \cos(\epsilon k_2
   + \tau \epsilon a_2)
   \sin(\epsilon k_1 + \tau \epsilon a_1)
   \right.
 \nonumber \\
 & & \left. - \, a_1 \cos(\epsilon k_1
   + \tau \epsilon a_1) \sin(\epsilon k_2 + \tau \epsilon a_2)
   \right]
 \nonumber \\
 & & \times \, \biggl \{ r^2 \biggl[ \sum_\mu ( \cos (\epsilon k_\mu
     + \tau \epsilon a_\mu) - 1) \biggr]^2
 \nonumber \\
 & &  + \, \sum_\mu  \sin^2 (\epsilon k_\mu
     + \tau \epsilon a_\mu)  \biggr \} ^{-1} \,,
 \end{eqnarray}
where $k$ runs over the momenta appropriate for our finite lattice.
In the limit $\epsilon \to 0$ one indeed recovers the expression
derived above by the point splitting technique in the continuum.
So our recipe for the calculation of $\mbox{Im} W$ from a lattice
gauge field configuration works at least in this (almost trivial)
case.

\section{OUTLOOK}

Obviously, there are many open problems which have to be solved
before our proposal can become useful. For example, how can we
deal with gauge field configurations leading to zero modes
of the Dirac operator $\DD (A_\tau)$ for some value of $\tau$?
According to ref.~\cite{alva}, for every pair of eigenvalues
crossing zero, $\mbox{Im} W$ picks up a contribution $\pm \pi$.
This problem has been ignored in our Schwinger model calculation,
so there might be additional contributions to $\mbox{Im} W$ for
certain configurations. If such configurations are statistically
relevant, $\exp (\mbox{i Im} W)$ would fluctuate strongly
and a Monte Carlo simulation might be hopeless, unless one
invents a clever method to update with a complex action. In any
case it would be interesting to see how other chiral fermion
proposals deal with this problem.

Even the calculation of $\mbox{Im} W$ for a single configuration
poses several numerical challenges. E.g.\ one has to compute
zero-mass fermion propagators in a given background field.
Furthermore, an operator trace has to be evaluated, which might
be feasible with the help of a stochastic estimator.

Let us close with the remark that the most elegant approach would
probably be to find a ``geometrical" expression for
$\mbox{Im} W$ based on the relation to the $\eta$-invariant.

\end{document}